\documentclass[letterpaper]{article}
\usepackage{aaai2026}

\usepackage{times}
\usepackage{helvet}
\usepackage{courier}
\usepackage[hyphens]{url}
\usepackage{graphicx}
\usepackage{natbib}
\usepackage{caption}
\usepackage{amsmath}
\usepackage{amssymb}
\usepackage{booktabs}
\usepackage{xcolor}
\usepackage{listings}
\usepackage[most]{tcolorbox}
\tcbuselibrary{listings,breakable}

\newtcblisting{promptbox}[2]{
  enhanced, breakable,
  colback=gray!5, colframe=gray!60!black, coltitle=white,
  fonttitle=\bfseries\small, title=#1,
  listing only,
  listing options={
    basicstyle=\ttfamily\scriptsize,
    breaklines=true,
    breakatwhitespace=false,
    columns=fullflexible,
  },
  boxrule=0.5pt, arc=1mm,
  left=2mm, right=2mm, top=1mm, bottom=1mm,
  #2
}

\title{Cyber Exodus: Burnout Symptoms, Exit Intention, and Peer Response in Online Cybersecurity Communities}

\author{
    Nadia Mehjabin\textsuperscript{\rm 1}, 
    Ji Hyun Kim\textsuperscript{\rm 1}, 
    Laura Barnes\textsuperscript{\rm 2}, 
    Koustuv Saha\textsuperscript{\rm 3}, 
    Henry Kautz\textsuperscript{\rm 1},  
    Subigya Nepal\textsuperscript{\rm 1}
}
\affiliations{
    \textsuperscript{\rm 1} Department of Computer Science, University of Virginia\\
    \textsuperscript{\rm 2} Department of Systems and Information Engineering, University of Virginia\\
      \textsuperscript{\rm 3} Siebel School of Computing and Data Science, University of Illinois Urbana-Champaign\\
    \ jqc7gj@virginia.edu, mqa4qu@virginia.edu, lb3dp@virginia.edu, ksaha2@illinois.edu, rmw7my@virginia.edu, sknepal@virginia.edu
}

\begin{document}

\maketitle

\begin{abstract}
Security practitioners burn out at high rates, and the resulting attrition is itself a security problem. This workforce is hard to study: security operations centers are closed to outside researchers, studies that reach practitioners recruit through employers, and those who have disengaged most may have the least reason to answer an employer's survey. The same practitioners discuss their working conditions openly in online communities. We adapt the Burnout Assessment Tool, a validated clinical instrument, into a text annotation scheme and apply it to 354,861 posts and 296,442 replies from five online communities of cybersecurity practitioners. Checked against two trained coders on 100 posts, the annotation reaches a macro F1 of 0.75 across the four symptoms and 0.98 for detecting any burnout signal. We find that the four symptoms point to different problems at work, not to the same problem at different levels of severity. Exhaustion appears in almost any complaint about staffing or workload. Mental distance, a loss of belief that the work is worthwhile, is the only symptom unrelated to operational problems, and among posts with a single symptom it is accompanied by a stated intention to leave roughly twice as often as any other. Peer responses show the opposite pattern. When a poster says they are considering leaving, the mix of replies shifts toward career advice, but this shift is smallest for mental distance. The symptom most strongly associated with leaving is thus the one peers adjust to least, and a single burnout score obscures both patterns.

\end{abstract}

\section{Introduction}
The cybersecurity workforce is under strain, and that strain is itself a security problem. In the 2025 ISC2 workforce survey, 48\% of practitioners reported exhaustion from keeping up with new threats and technologies, 47\% reported being overwhelmed by their workload, and a third said their organization lacked the resources to staff its team \citep{isc22025workforce}. More than half of organizations report difficulty retaining qualified security staff \citep{isaca2024state}. The consequences are operational. Sustained alert volume desensitizes analysts and degrades their effectiveness, which recent work identifies as an open challenge for the field \citep{tariq2025alert}, and organizations with severe security staffing shortages incur breach costs averaging 1.76 million US dollars more than those with little or no shortage \citep{ibm2024breach}. When experienced analysts leave, those who remain absorb the work, which can raise fatigue, let detection slip and prompt further departures.

Despite this, the human side of security is harder to study than the technical side, and the obstacle is access. Security operations centers are closed environments, and empirical work inside them is routinely blocked by non-disclosure agreements, liability concerns and operational security requirements. Research that does reach this workforce recruits through employers, professional networks or targeted outreach \citep{hollis2023all,arora2024survey,nepal2024burnout}. That work establishes that the problem exists, but it does not scale, and it misses a particular group. Security work carries clearance requirements and strong norms around anonymity, and admitting to burnout can read as admitting to reduced reliability. In one survey, 60\% of practitioners said they were unlikely to report work-related stress or burnout to management \citep{arora2024survey}, and workers more generally fear that wellbeing data will be used to evaluate them \citep{kawakami2023wellbeing}. The people furthest along, who have stopped caring and are preparing to leave, may have the least reason to answer.

Online communities of practitioners offer a way around that barrier. They work as informal break rooms in which practitioners discuss conditions they cannot raise at work. The disclosure is unsolicited rather than prompted by a researcher, and every reply is preserved beside it. Prior work has established this kind of text as a source for studying distress and the support that follows it \citep{de2014mental,saha2020causal,alghamdi2025redditess}, mostly in general mental health communities rather than in a single occupation.

Two questions remain open for this workforce. The first concerns the composition of its burnout. The Maslach Burnout Inventory and the Burnout Assessment Tool both separate burnout into distinct symptoms \citep{maslach1981measurement,schaufeli2020burnout}. Studies of security practitioners, however, mostly report burnout as a single figure or describe it qualitatively \citep{arora2024survey,pham2019information,hollis2023all}, and where the symptoms have been measured separately in a recruited sample \citep{nepal2024burnout} they have not been tied to the working situations practitioners describe or to their intention to leave. So for this workforce it is not known whether a practitioner reporting exhaustion and one reporting cynicism are describing the same conditions or carry the same risk of departure. If they are not, a workforce-level burnout figure would conceal which practitioners are close to leaving. The second question is what happens after someone discloses. A survey can record that a practitioner is burned out. It cannot record what their peers said in reply, and for many practitioners peers are the main support available.

We therefore ask: \textit{How do the four burnout symptoms relate to the work situations and intentions to leave that cybersecurity practitioners express in online posts, and how do peer responses vary across these symptoms?}

To address this question, our contributions are as follows:

\begin{enumerate}
\item \textbf{The symptom closest to leaving is the one peers answer least.} Among posts carrying a single symptom, mental distance co-occurs with a stated intention to leave roughly twice as often as any other symptom. It is also the symptom whose replies shift least when that intention is stated, and the ordering holds for every reply type we can compare (Sections~\ref{sec:exit} and~\ref{sec:comment-analysis}). Neither community membership nor voting behavior accounts for this.

\item \textbf{The four symptoms describe different situations, not one condition at four strengths.} Rather than relying on the symptom definitions, we let a concept induction method read the posts and report what they are about (Section~\ref{sec:concepts}). Exhaustion accompanies almost any complaint about staffing and workload. Mental distance is the only symptom with no detected association with operational content, appearing instead with technical disillusionment and mismatched values.

\item \textbf{Construct-level burnout measurement in unsolicited text.} We adapt the Burnout Assessment Tool into an annotation scheme and apply it to 354,861 posts spanning 2018 to 2026 (Sections~\ref{sec:corpus}--\ref{sec:bat}). We validate the symptom and exit labels against two trained coders, and the symptoms against a model that estimates how much each rater catches without treating any rater as ground truth (Section~\ref{sec:validity}).
\end{enumerate}

\section{Related Work}

\subsection{Burnout in Security Work}
Studies of security practitioners converge on the same picture from several directions. \citet{hollis2023all} ties stress and burnout among incident responders to multifaceted job demands, improvisation and emotional labour. \citet{arora2024survey} find that 44\% of 50 practitioners report severe work-related stress and burnout, and most often attribute it to the nature of the work, unsupportive cultures and unrealistic expectations. \citet{nepal2024burnout} link burnout among incident responders to job demands and poor sleep, and identify the unpredictable timing of incidents as a leading stressor. An ethnography of a corporate security operations center describes analyst burnout as the product of human, technical and managerial factors interacting over time \citep{sundaramurthy2015human}. Related work on employees outside the security function ties burnout to the demands of complying with security policy \citep{pham2019information} and shows that work overload raises job stress, which in turn degrades security behavior \citep{hong2023mitigating}.

These studies span several roles and countries. Each recruits through channels that require institutional access, so none can scale, and none observes what follows a disclosure. Where they establish how much burnout exists in recruited samples, we ask what its components look like in text practitioners wrote for their peers, and what those peers wrote back.

\subsection{Distress and Support in Online Communities}
A large body of work uses public online text to study how people disclose difficult experiences and what support they receive \citep{de2014mental,andalibi2018social}. The language of replies predicts later outcomes for the person who posted \citep{saha2020causal}, supporters differ in whether they offer information or emotional care \citep{kim2023supporters}, and reciprocal exchanges keep people in a community \citep{sharma2020engagement}. \citet{alghamdi2025redditess} go further and ask whether a supportive reply is effective rather than merely present, using community signals such as votes alongside the text and grounding their labels in social science theories of support.

\citet{chancellor2020methods} reviewed methods for predicting mental health status from social media and found recurring problems with how constructs are defined and how validity is established. \citet{ernala2019methodological} showed that classifiers trained on social media proxies for a diagnosis performed poorly on patients with clinically verified diagnoses. We take two lessons from that work. We do not predict a diagnosis, only the presence of symptom criteria in a piece of text, and we report what our labels are worth against human coders rather than treating model output as ground truth.

Further, our framing is occupational rather than clinical: our posters describe a job, and the risk they take in disclosing is a career risk rather than a social one. And we condition support on the symptom profile of the post being replied to, rather than treating support as a property of the community in general. That lets us ask whether the help offered matches the difficulty disclosed, which is the question optimal matching theory poses for support more broadly \citep{cutrona1990type}.

\subsection{Measuring Burnout and Intention to Leave}
\label{sec:relwork-exit}

Burnout is commonly measured through self-report instruments such as the Maslach Burnout Inventory~\citep{maslach1981measurement} and the Burnout Assessment Tool (BAT)~\citep{schaufeli2020burnout}. The BAT has been validated across occupations and countries~\citep{schaufeli2020burnout,angelini2021burnout}, and its 12-item short form converges with the Maslach instrument~\citep{de2022investigating}. Social media research has also operationalized distinct facets of job satisfaction from posts about work~\citep{saha2021social}.

Intentions to leave are associated with subsequent departure~\citep{hom2017one}, and fit Hirschman's account of exit as one response to a deteriorating organization~\citep{hirschman1970exit}. Studies of online communities have similarly examined self-disclosed accounts of disengagement from social and ideological groups~\citep{phadke2025exit}. Research on occupational turnover identifies working conditions, workplace relationships, recognition, and health among the reasons employees give for leaving~\citep{horberg2023experienced}. Person--organization fit research also links alignment between individual and organizational values to work attitudes and withdrawal~\citep{kristof2005consequences,cable1996person}.

Applying these constructs to online text requires explicit definitions and evidence about what the resulting measures capture~\citep{chancellor2020methods,ernala2019methodological}. Studies of language models demonstrate their potential for text annotation~\citep{gilardi2023chatgpt,tornberg2023chatgpt}, with performance varying across tasks and constructs~\citep{ziems2024can}. We build on this work by adapting BAT criteria and turnover categories to identify burnout symptoms, intentions to leave, and reasons for departure in practitioner posts. We evaluate the burnout labels against human coding and assess agreement across models for the remaining annotations.

\section{Data and Methods}
Our analysis proceeded in five stages. After constructing the corpus, we screened posts for work-related distress, annotated burnout symptoms and intentions to leave, characterized the content of posts and replies, and examined temporal associations between vulnerability disclosures and burnout-related posting.

\subsection{Corpus}
\label{sec:corpus}

\begin{figure*}[t]
\centering
\includegraphics[width=0.93\textwidth]{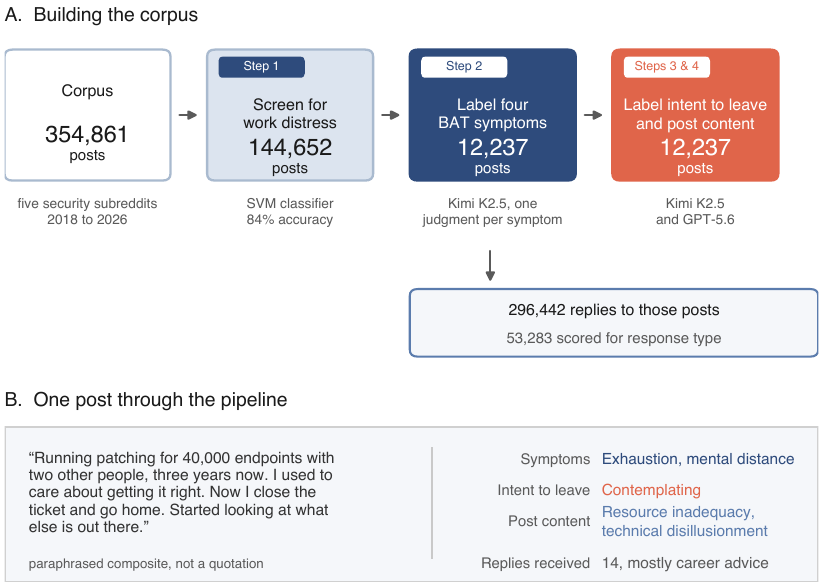}
\caption{Panel A: how many posts remain after each stage, and what was used to get there. The 296,442 replies we analyze are the replies to the 12,237 posts that carry at least one symptom. Panel B: a single post, and everything the pipeline records about it.}
\label{fig:pipeline}
\end{figure*}

We collect posts from r/sysadmin, r/cybersecurity, r/asknetsec, r/SecurityCareerAdvice and r/ciso using Arctic Shift \citep{arcticshift}, an archival service that provides bulk access to historical Reddit data. We removed duplicate submissions, kept only self-posts containing user-written text, restricted the window to January 2018 through April 2026, dropped posts whose body had been deleted, stripped URLs and markdown, and excluded posts under 20 words. Minimum-length thresholds in this literature vary by task; \citet{hengle2024still}, for example, drop Reddit posts under 75 words. This yielded 354,861 posts. Table~\ref{tab:corpus} gives the breakdown by community, and Figure~\ref{fig:pipeline} shows how the corpus narrows at each step of the pipeline described below.

\begin{table}[t]
\centering
\small
\setlength{\tabcolsep}{4pt}
\caption{The corpus by community. Symptomatic posts show at least one BAT symptom. Rate is the share of that community's posts that are symptomatic, which orders the communities differently from the raw counts.}
\label{tab:corpus}
\begin{tabular}{lrrr}
\toprule
Community & Posts & Sympt. & Rate \\
\midrule
r/sysadmin             & 269{,}343 & 9{,}871 & 3.7\% \\
r/cybersecurity        & 61{,}586  & 1{,}626 & 2.6\% \\
r/asknetsec            & 15{,}832  & 199     & 1.3\% \\
r/SecurityCareerAdvice & 7{,}674   & 529     & 6.9\% \\
r/ciso                 & 426       & 12      & 2.8\% \\
\midrule
Total                  & 354{,}861 & 12{,}237 & 3.4\% \\
\bottomrule
\end{tabular}
\end{table}

We treat all five as security practitioner communities. r/sysadmin is the largest, and we include it because the security function in most organizations is not staffed separately from infrastructure operations. Outside large enterprises, the same people run patching, identity, logging, and incident response, and the conditions most tied to burnout in this work fall on that same group. Excluding them would define the profession by job title rather than by the work performed. We report r/sysadmin as its own stratum throughout and test explicitly whether it behaves differently from the four communities that name security in their titles.

\subsection{Step 1: Identifying Work-Related Distress}
\label{sec:screen}

We developed a screening scheme covering workload, work-life balance, imposter syndrome, job search difficulties, and motivation. The scheme included descriptions of distress that did not explicitly mention burnout. We manually labeled 410 posts using these criteria; Table~\ref{tab:screening} and Appendix~\ref{app:screen} provide the annotation scheme.

We evaluated three screening approaches on the same 82-post holdout set. A few-shot Kimi K2.5 prompt achieved 87.8\% accuracy, a linear support vector machine (SVM) using word-frequency features achieved 84.2\%, and a zero-shot Kimi K2.5 prompt achieved 82.9\%. We selected the SVM because its accuracy was within 3.6 percentage points of the best-performing approach and it could screen the full corpus without a language-model call for each post. The SVM identified 144,652 posts (40.8\%) as expressing work-related distress. These posts proceeded to the symptom and exit-intention annotation stages.

\subsection{Step 2: Annotating Burnout Symptoms}
\label{sec:bat}

We operationalized burnout expressions using the four core symptoms of the Burnout Assessment Tool (BAT)~\citep{schaufeli2020burnout}. \emph{Exhaustion} concerns physical or mental depletion. \emph{Emotional impairment} concerns intense or persistent work-related emotional reactions. \emph{Cognitive impairment} concerns difficulties with memory, attention, or decision-making at work. \emph{Mental distance} concerns persistent psychological withdrawal from work, including indifference, cynicism, or working on autopilot.

We adapted items from the validated BAT short form into textual inclusion and exclusion criteria for each symptom~\citep{de2022investigating}. Kimi K2.5 assessed each screened post separately for each symptom and returned a supporting text span for every positive label. Posts could receive multiple labels. We summed the four binary labels to obtain a symptom count ranging from 0 to 4. Appendix~\ref{app:bat} provides the annotation prompt.

\subsection{Validation of Burnout Labels}
\label{sec:validity}

Two trained coders annotated 100 posts: 40 selected at random and 60 sampled to provide roughly equal coverage of the four symptoms. We measured coder agreement using Gwet's AC1, a chance-corrected coefficient that addresses the prevalence-related agreement paradox associated with Cohen's $\kappa$~\citep{gwet2008computing}.

Coder agreement was 0.93 for the presence of any burnout signal and 0.63 across the four symptom labels (Table~\ref{tab:validity}). Agreement varied by symptom: 0.84 for exhaustion, 0.73 for cognitive impairment, 0.52 for emotional impairment, and 0.54 for mental distance. These differences indicate greater consistency in identifying exhaustion and cognitive impairment than in distinguishing emotional impairment and mental distance.

We evaluated model performance on the cases where the two coders agreed. For any burnout signal, the model achieved an F1 of 0.98 and 97\% agreement with the shared coder label. Across the four symptoms, macro F1 was 0.75. F1 ranged from 0.81 to 0.84 for exhaustion, emotional impairment, and mental distance. For mental distance, the model achieved an F1 of 0.83 and balanced accuracy of 0.87 on the 76 coder-agreed posts.

Cognitive impairment had a lower F1 of 0.52. Among the 81 coder-agreed posts, eight were positive; the model identified six of these and labeled nine coder-negative posts as positive. Its balanced accuracy was 0.81. We report F1 alongside balanced accuracy to show performance on positive cases and across both classes.

\begin{table}[t]
\centering
\small
\setlength{\tabcolsep}{3.5pt}
\caption{Coder agreement and model performance, on 100 validation posts for the symptoms and a separate 100 for exit intention. $n$ is the number of posts where both coders agreed and \%Y the share of those that were positive. The model is scored against the coder-agreed label.}
\label{tab:validity}
\begin{tabular}{lrrrrr}
\toprule
Symptom & $n$ & \%Y & AC1 & F1 & Bal. \\
\midrule
Exhaustion            & 91 & 67 & 0.84 & 0.84 & 0.82 \\
Emotional impairment  & 76 & 50 & 0.52 & 0.81 & 0.80 \\
Cognitive impairment  & 81 & 10 & 0.73 & 0.52 & 0.81 \\
Mental distance       & 76 & 38 & 0.54 & 0.83 & 0.87 \\
\midrule
Exit intention, three classes & 85 & 53 & 0.78 & 0.96 & 0.97 \\
Any exit intention    & 88 & 55 & 0.76 & 0.98 & 0.98 \\
\midrule
All four symptoms     & 324 & 42 & 0.63 & 0.75 & 0.83 \\
Any burnout signal    & 95 & 81 & 0.93 & 0.98 & n/a \\
\bottomrule
\end{tabular}
\end{table}

We also fitted a latent class model to the three raters' annotations on
all 100 posts, including the cases where the coders disagreed. The model
jointly estimated latent symptom status and each rater's sensitivity and
specificity from their annotation patterns (Table~\ref{tab:latentclass}).
The model is the weakest of the three raters on exhaustion. It
over-reports emotional impairment, passing over fewer non-cases than
either coder. It is the strongest rater on cognitive impairment, the
symptom the coders found hardest to agree on. On mental distance the two
coders applied divergent thresholds, one conservative and one liberal,
and the model fell between them on both measures, so the low coder
agreement for that symptom reflects a difference in coder threshold
rather than a symptom the model cannot detect. These model-based
estimates complement the performance measures calculated on coder-agreed
cases.

\subsection{Step 3: Annotating Intention to Leave}
\label{sec:exitmethod}

We classified each screened post into one of three categories. \emph{Explicit} indicated a stated intention to leave a job or the profession, such as giving notice. \emph{Contemplating} indicated consideration of leaving, including tentative plans to explore other opportunities. \emph{No exit} indicated no expressed intention to leave. Kimi K2.5 assigned these labels in a separate annotation pass and returned a confidence score, supporting text span, and short justification. The criteria concerned language about leaving and did not refer to burnout symptoms. Appendix~\ref{app:exit} provides the prompt.

The same two coders evaluated a separate set of 100 posts sampled to include roughly equal numbers of the three categories. Coder agreement was 0.78 for the three-class annotation, and the model achieved a macro F1 of 0.96 on the 85 coder-agreed posts. When explicit and contemplating labels were combined into a binary exit-intention measure, coder agreement was 0.76. On the 88 posts where coders agreed on this binary label, the model achieved an F1 of 0.98, identified all coder-positive cases, and labeled two coder-negative cases as positive. Figure~\ref{fig:inversion} uses this binary measure.

\subsection{Step 4: Characterizing Posts and Replies}
\label{sec:concepts}
We used LLooM~\citep{lam2024concept} with \texttt{gpt-5.6-luna-pro} to generate candidate concepts from sampled documents. LLooM distills documents into short descriptions, groups related descriptions, and generates concept names and inclusion criteria. This process allowed us to identify recurring content without supplying a predefined list of content categories.

\paragraph{Categories from prior research.}
For posts, we adapted the turnover categories discussed in Section~\ref{sec:relwork-exit}. For replies, we used five categories of social support~\citep{cutrona1992controllability} and a category for unsupportive responses~\citep{ingram2001unsupportive}. These categories connected the analysis to established accounts of turnover and social support. We finalized the induced concepts before writing the criteria for categories drawn from prior work.

\paragraph{Post content.}
We ran concept induction on 2,000 of the 12,237 symptomatic posts, sampling roughly equal numbers from each observed symptom combination to represent less common combinations. Although we initially targeted approximately 20 concepts, four induced concepts were retained after review. Together with six categories adapted from prior work, these formed the final set of ten post-content labels. Table~\ref{tab:labels} in Appendix~\ref{app:labels} lists the labels and their definitions.

\paragraph{Reply content.}
The reply corpus contained 296,442 top-level comments on symptomatic posts. Approximately 90\% came from r/sysadmin. We annotated all comments from the four smaller communities and a random sample of 25,000 from r/sysadmin, yielding 53,283 comments. For corpus-level estimates, we weighted the sampled r/sysadmin comments to recover that community's share of the full reply corpus.

We directed reply induction toward the functions of responses, such as offering guidance or sharing an experience. We selected a steering phrase by comparing how many concepts generated by candidate phrases described response functions. Eight induced concepts were retained after review: career planning advice, workplace problem guidance, practical advice, emotional support, personal relating, support connections, critical pushback, and discussion direction. Appendix~\ref{app:labels} provides their criteria.

\paragraph{Label agreement and stability.}
We assessed agreement across three models from different providers: \texttt{gpt-5.6-luna-pro}, \texttt{claude-sonnet-4.5}, and \texttt{gemini-2.5-pro}. Each model annotated the same 300 documents, and we calculated Gwet's AC1 for each label. Ten of the twelve candidate post labels exceeded the 0.60 agreement threshold, with AC1 values ranging from 0.70 to 0.94. We retained these ten labels and excluded personal limits (0.52) and systemic futility (0.55).

We also assessed label stability by rewording the criteria and reversing document order. These changes altered an average of 6.3\% of labels. Personal limits and systemic futility were the least stable, with changes of 15.4\% and 9.4\%, respectively. Using the retained criteria, \texttt{gpt-5.6-luna-pro} annotated all 12,237 symptomatic posts and the 53,283 sampled comments. For analyses of replies, we clustered standard errors by post to account for dependence among comments responding to the same post.

\subsection{Step 5: Temporal Associations with Vulnerability Disclosures}
\label{sec:granger}
Security work has obvious external shocks. A serious software vulnerability becomes public, and for the following weeks practitioners patch systems under time pressure. If burnout in these communities is driven by those events, then posting about burnout should increase in the weeks after a spike in serious vulnerabilities. If, instead, it accumulates gradually, no such link should appear.

We tested this against the National Vulnerability Database (NVD), maintained by the US National Institute of Standards and Technology~\citep{nistnvd}. We constructed weekly counts of disclosed vulnerabilities with severity scores above 9.5 on the 0--10 scale and compared them with weekly counts of symptomatic posts. We applied Granger causality tests to assess whether past vulnerability counts improved prediction of symptomatic posting beyond the posting series' own history~\citep{granger1969investigating}. We adjusted for comparisons across multiple time lags using Holm's correction~\citep{holm1979simple}.

\section{Results}

\subsection{Burnout Expressions Across Communities}

Of the 144,652 posts retained by the screening stage, 12,237 (8.5\%) received at least one burnout symptom label. The number of posts decreased with symptom count, from 7,232 with one symptom to 389 with all four. Emotional impairment was the most common label, appearing in 5.3\% of screened posts, followed by exhaustion (4.3\%), mental distance (2.3\%), and cognitive impairment (1.6\%).

Community-level rates, calculated using all retained posts in each community, are reported in Table~\ref{tab:corpus}. r/sysadmin contributed the largest number of symptomatic posts. The proportion of posts labeled symptomatic was highest in r/SecurityCareerAdvice (6.9\%) and lowest in r/asknetsec (1.3\%). We report r/ciso descriptively, given its 12 symptomatic posts.

\subsection{Work Situations Associated with Burnout Symptoms}

\begin{figure*}[t]
\centering
\includegraphics[width=0.95\textwidth]{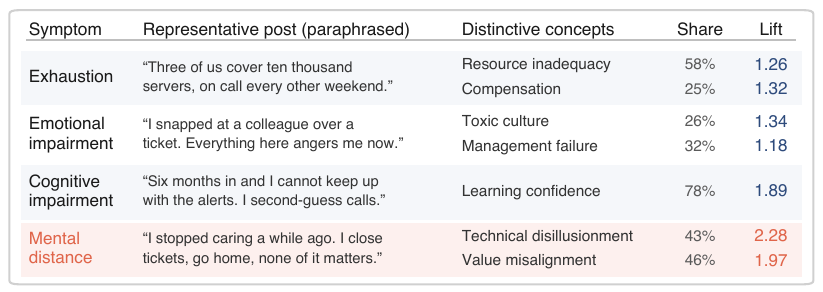}
\caption{Work situations associated with each burnout symptom. Share indicates the percentage of posts expressing a given symptom that also contain the content label. Lift is how many times more often it appears there than in posts without it, so a lift of 2 means twice as common. Exhaustion sits close to 1 on everything, while mental distance reaches 2.28. Examples are paraphrased composites.}
\label{fig:card}
\end{figure*}

Within the symptomatic corpus, we compared the prevalence of each content label in posts with and without a given symptom. Figure~\ref{fig:card} summarizes the content profiles.

\paragraph{Exhaustion.}
Nine of the ten content labels were significantly associated with exhaustion, although the associations were modest. These included compensation, resource inadequacy, and workplace career strain. Exhaustion thus appeared across several forms of work strain.

\paragraph{Emotional impairment.}
Toxic culture appeared in approximately a quarter of posts expressing emotional impairment, compared with 8.3\% of posts without this symptom. Management failure and security governance resistance were also associated with emotional impairment. Workload and compensation showed no statistically significant association.

\paragraph{Cognitive impairment.}
Learning confidence struggles were the most common content label, appearing in 78.2\% of posts expressing cognitive impairment and 32.6\% of posts without it. Value misalignment, toxic culture, and compensation were significantly less common in posts expressing cognitive impairment.

\paragraph{Mental distance.}
Technical disillusionment appeared in 42.7\% of posts expressing mental distance, compared with 9.5\% of posts without it. Value misalignment appeared in 45.6\% and 14.5\%, respectively. Resource inadequacy and IT operations dysfunction showed no statistically significant association with mental distance; both were associated with exhaustion.

Concept rankings were broadly similar across symptom-count groups (Spearman's $\rho = 0.867$). Technical disillusionment was present in 13.2\% of posts with one symptom and 44.7\% of posts with all four, indicating a higher prevalence in posts expressing multiple burnout symptoms.

\subsection{Burnout Symptoms and Intention to Leave}
\label{sec:exit}

Across the screened corpus, 3.70\% of posts were labeled as contemplating leaving, and 1.05\% expressed an explicit intention to leave. The proportion expressing either form of exit intention increased with symptom count, from 2.7\% of posts with no symptoms to 51.3\% of posts with all four. Table~\ref{tab:exitex} illustrates the three exit-intention categories.

\begin{table}[t]
\centering
\small
\setlength{\tabcolsep}{4pt}
\caption{Illustrative examples of the exit-intention categories, presented as paraphrased composites. Categories are defined by language about leaving.}
\label{tab:exitex}
\begin{tabular}{@{}lp{5.4cm}@{}}
\toprule
Class & Representative post \\
\midrule
No exit & ``Three of us cover ten thousand servers
and I am on call every other weekend.'' \\
\addlinespace[2pt]
Contemplating & ``I have started looking at what
else is out there, though I am not sure I want to
leave the field.'' \\
\addlinespace[2pt]
Explicit & ``I put my notice in last week. Six
years of this was enough.'' \\
\bottomrule
\end{tabular}
\end{table}

Exit intention also varied by symptom. Approximately half of posts expressing mental distance contained an intention to leave, compared with roughly a fifth of posts expressing cognitive impairment. To examine this pattern without symptom co-occurrence, we restricted the comparison to posts expressing exactly one symptom. Exit intention appeared in 40.7\% of posts expressing mental distance, compared with 20.8\% for exhaustion, 11.7\% for emotional impairment, and 7.9\% for cognitive impairment. The rate for mental distance was nearly twice that for exhaustion and higher still relative to the other two symptoms.

Post content differed between contemplating and explicit exit intentions. Toxic culture appeared in 25.6\% of posts contemplating departure and 42.5\% of posts expressing an explicit intention to leave. Compensation, management failure, and resource inadequacy were also more common in the explicit category. Learning confidence struggles showed the opposite pattern, appearing in 46.1\% of contemplating posts and 25.2\% of explicit posts.

\paragraph{Temporal associations with vulnerability disclosures.}
Most tests relating severe vulnerability disclosures to subsequent burnout-related posting or exit intention were not statistically significant after Holm correction. One association remained significant: higher counts of severe vulnerability disclosures preceded an increase in posts expressing at least two burnout symptoms at a four-week lag ($p = .004$, $p_{\text{Holm}} = .043$). We do not read this as a relationship. It is a single pairing of one severity cutoff with one symptom threshold at one lag, it sits just inside the corrected boundary, and it is roughly what testing this many combinations would produce by chance. The test also compares one week against the next, so it cannot detect a process that accumulates over months, which is both what burnout is by definition and what our severity results point toward.

\subsection{How Peers Respond}
\label{sec:comment-analysis}

\begin{figure*}[t]
\centering
\includegraphics[width=0.90\textwidth]{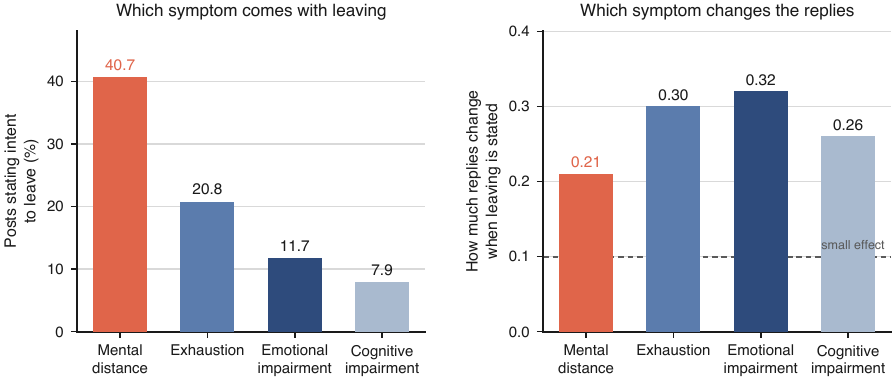}
\caption{Exit intention and peer responses by burnout symptom. Left: the percentage of posts expressing a single symptom that also express an intention to leave. Right: the strength of association between exit intention and career planning advice within each symptom group. Larger values indicate a stronger association. Mental distance has the highest exit-intention rate and the weakest association between exit intention and reply type. This ordering also holds for the two other reply types observed across all four symptoms.}
\label{fig:inversion}
\end{figure*}

\begin{table}[t]
\centering
\small
\caption{Common reply types, their corresponding social-support categories, and paraphrased composite examples.}
\label{tab:replycard}
\begin{tabular}{@{}p{\dimexpr\columnwidth-2\tabcolsep\relax}@{}}
\toprule
\textbf{Career planning advice} \hfill \textit{informational} \\
``Move into GRC and you are off the pager.'' \\
\addlinespace[3pt]
\textbf{Workplace problem guidance} \hfill \textit{informational} \\
``Get every override in writing and copy your manager.'' \\
\addlinespace[3pt]
\textbf{Personal relating} \hfill \textit{network} \\
``I did four years in a SOC and felt exactly this.'' \\
\addlinespace[3pt]
\textbf{Emotional support} \hfill \textit{emotional, esteem} \\
``That sounds exhausting, and none of it is your fault.'' \\
\addlinespace[3pt]
\textbf{Critical pushback} \hfill \textit{unsupportive} \\
``This is just what the job is.'' \\
\bottomrule
\end{tabular}
\end{table}

Advice was prominent in the replies, with career planning advice the most common reply type. Table~\ref{tab:replycard} illustrates five frequently occurring types, including workplace guidance, shared experiences, emotional support, and critical pushback. Reply patterns varied across exit-intention categories, with the strongest association for career planning advice and the weakest for critical pushback.

We quantified the association between exit intention and reply type within each burnout symptom group using Cram\'er's $V$. Figure~\ref{fig:inversion} presents this comparison for career planning advice alongside the single-symptom exit-intention rates. The association was strongest for emotional impairment and exhaustion, followed by cognitive impairment, and weakest for mental distance. This pattern held for all three reply types observed across the four symptoms. The association for mental distance was $V = 0.21$.

We repeated the analysis on posts expressing exactly one symptom, and the same pattern held. Mental distance therefore combined the highest frequency of stated exit intention with the weakest association between exit intention and peer response type.

\subsection{Community Differences and Voting Patterns}
\label{sec:confounds}
We considered two explanations for this pattern that have nothing to do with the symptom itself.

The first is that different communities might simply reply in different ways, and mental distance posts might be concentrated in a community with an unusual reply style. Comparing r/sysadmin against the four security-focused subreddits, only 4 of 14 reply types differ meaningfully between them, and all four are kinds of advice. That is a difference in what people ask about, not in how they support each other.

The second is that these communities might already be correcting the pattern through voting, rewarding better replies over time. They are not. Whether a comment is upvoted, and how highly, has essentially no relationship to what kind of reply it is. No style of response is rewarded or penalized, so nothing here would push these communities toward responding differently.

\section{Discussion}
We asked what could be learned from what security practitioners already write in public, and found that the answer is not simply a larger version of what surveys report.

\paragraph{Burnout symptoms and working conditions.}
The content associated with each symptom helps identify what practitioners find difficult about their work. Exhaustion appeared across several forms of work strain, consistent with the Job Demands--Resources account of demands and exhaustion~\citep{demerouti2001job}. Emotional impairment co-occurred with toxic culture and management failure, cognitive impairment with learning confidence struggles, and mental distance with technical disillusionment and value misalignment. These patterns extend research on burnout among incident responders~\citep{nepal2024burnout} by connecting burnout symptoms to the situations practitioners describe in peer discussions.

The differences also matter for interpreting departure intentions. Among posts expressing a single symptom, 40.7\% of those expressing mental distance mentioned leaving, compared with 20.8\% for exhaustion. Reporting individual symptoms alongside an overall burnout score would make these differences visible. In workplace assessments, this could guide questions about workload, management relationships, confidence in meeting job demands, and alignment with organizational values.

\paragraph{Public text reaches what a survey cannot.} There is a reason the symptom closest to leaving is also the one least likely to appear on a form. Telling an employer you are exhausted reads as someone working hard under strain, and it invites help. Telling an employer you no longer care about the work can raise concerns about your commitment and reliability. Practitioners who are considering leaving therefore have stronger reasons to stay quiet.

Public communities reduce this cost. The disclosure is unsolicited rather than prompted by a researcher, it is made to peers rather than to anyone with authority over the writer, and it is made under a pseudonym. We cannot show that practitioners answer surveys differently, because we did not measure that. We offer it as the most plausible reason that these two sources would see different things.

\paragraph{Peer support and the difficulty disclosed.}
Optimal matching theory proposes that the usefulness of support depends on its fit with the stressor~\citep{cutrona1990type}. Research in online mental health communities has similarly connected the language and adaptability of peer responses to psychosocial outcomes~\citep{saha2020causal}, and examined how supporters differ in the informational and emotional resources they provide~\citep{kim2023supporters}. Our findings bring the form of occupational distress into this account: the association between exit intention and reply type varied across burnout symptoms.

One possible explanation concerns how readily a post lends itself to advice. Staffing problems or career transitions may provide familiar openings for practical suggestions. Disillusionment and value conflicts may require more information about what the practitioner wants from the exchange. This interpretation motivates examining the fit among the difficulty disclosed, the response offered, and the recipient's assessment of its usefulness.

\paragraph{Learning from occupational disclosures.}
Online communities make it possible to examine accounts of work alongside the interactions that follow them. Prior research has linked anonymity with disclosure and support seeking~\citep{de2014mental}, while studies of workplace well-being technologies document concerns about how employers might interpret and use well-being information~\citep{kawakami2023wellbeing}. Pseudonymous communities may provide opportunities to discuss disengagement outside formal workplace communication.

Our approach also extends research that operationalizes workplace experiences through social media~\citep{saha2021social}. Adapting BAT items into explicit annotation criteria allowed us to examine specific burnout expressions and connect them to peer responses. Applying this approach to other occupational communities would require attention to how the relevant constructs are expressed and validated in each setting~\citep{chancellor2020methods,ernala2019methodological}.

\paragraph{Implications for community support.}
The prominence of advice suggests opportunities to help practitioners communicate what they want from a response. Optional posting prompts could invite requests for practical guidance, shared experiences, or emotional support. Guidance for responders could include acknowledging disillusionment, asking what kind of help would be useful, and sharing relevant experiences. These approaches could be developed with community members and evaluated through recipients' judgments of helpfulness and supporters' experiences of responding.

\paragraph{Preserving practitioner control.}
Research and practical applications should preserve practitioners' control over how their disclosures are used. Studies of workplace well-being technologies show how organizational power can complicate consent and create conflicting expectations about data use~\citep{kawakami2023wellbeing,chowdhary2023can}. Applications should support voluntary participation, confidentiality, and reporting at the community level. Linking pseudonymous disclosures to individual employment decisions would undermine this control. Community members should also have a role in defining useful research questions and how findings are communicated.

\paragraph{What we are not proposing.} We do not suggest that employers run this measurement on their staff or that these communities be monitored. Practitioners write in them because they sit outside the employment relationship, and detection by an employer would remove the condition that makes the writing possible. The value of this corpus for research depends on it not being used that way.

\section{Limitations and Future Directions}

Our study has limitations that suggest several directions for future research.

\paragraph{Sampling and disclosure.}
The corpus reflects self-selected participation in five Reddit communities, with r/sysadmin contributing most posts and replies. Our estimates describe expressions within this corpus and depend on what practitioners choose to disclose. Future studies could examine additional communities and use voluntary recruitment to compare online accounts with confidential surveys and interviews. Such comparisons could clarify which experiences are disclosed in different settings and how the observed patterns vary across occupational roles.

\paragraph{Measurement and classification.}
Errors in screening and symptom annotation can affect subsequent comparisons. Coder agreement was lower for emotional impairment and mental distance, and model F1 was lowest for cognitive impairment. The latent class analysis also estimated lower model sensitivity for exhaustion than for either human coder. Larger validation samples stratified by symptom, community, and year, including posts excluded by screening, could better characterize these errors and assess how they affect the findings. Pairing text annotations with consenting practitioners' BAT responses could further examine the relationship between expressed symptoms and self-reported burnout.

\paragraph{Content and response categories.}
We assessed content and reply labels through agreement across three models, which may share training data or interpretive assumptions. Rewording criteria and changing document order altered an average of 6.3\% of labels. The four retained induced post concepts also reflect the sampling and concept-generation process. Future work could combine human annotation with inductive qualitative analysis across symptom groups to identify shared model errors and experiences missing from the current categories.

\paragraph{Peer responses and their usefulness.}
Our analysis covers top-level replies; support developed in nested conversations, private messages, or offline interactions falls outside this view. The reply categories and association measures describe response patterns, and we did not measure recipients' assessments of helpfulness. Extending the analysis to complete threads and voluntary recipient feedback could clarify how support develops over an exchange, building on research on reciprocal interaction and community participation~\citep{sharma2020engagement}. Comparisons of posts describing similar work situations, together with vignette studies that vary symptom and exit language, could also help distinguish how each contributes to the responses offered.

\paragraph{Temporal patterns and departure.}
Our main comparisons are at the post level, and the temporal analysis uses weekly vulnerability disclosures and aggregate posting counts. These measures provide limited information about individual exposure to security incidents or changes in working conditions. Longitudinal studies with consenting practitioners could connect repeated burnout measures with work demands, incident exposure, and employment changes. This would help distinguish co-occurring symptoms from changes over time and examine how stated intentions to leave relate to subsequent departure.

\section{Conclusion}
This study examines burnout expressions, departure intentions, and peer responses in cybersecurity communities. The four burnout symptoms were associated with different work concerns. Mental distance co-occurred most frequently with intentions to leave and showed the weakest association between exit intention and reply type. These findings show the value of examining specific burnout expressions together with the interactions that follow their disclosure. They also motivate research on the forms of support practitioners find useful when describing psychological withdrawal from work.

\section*{Acknowledgements}

Generative AI tools were used to support the development of code and figures and the editing of prose. Their use as research instruments, performing the annotation described in Steps 1 to 4, is reported in the Methods and validated in Sections~\ref{sec:validity} and~\ref{sec:concepts}. The authors designed the study, performed all analyses, interpreted results and prepared the final text.

\bibliography{aaai2026}

\section*{Paper Checklist}

\subsection*{Ethical Statement}

All data analyzed in this study comes from Reddit's publicly accessible submissions and comments, retrieved through an archival service in accordance with the platform's terms. We report analyses exclusively at the aggregate level. No usernames or user identifiers are released, and usernames were hashed after being used to filter bot accounts. We construct no user-level profiles and track no individual across posts.

Every post excerpt in this paper is a paraphrased composite written from the flagged material rather than a quotation, because verbatim Reddit text can be searched back to the account that produced it. Post identifiers are sequential rather than real. We excluded r/ciso from all group comparisons, because with 12 symptomatic posts an individual could plausibly be identified from a reported cell.

This work infers a mental health construct about people who did not consent to be studied. We therefore report population-level patterns only. This work is not intended to support identification of individuals, and we do not recommend its use inside an employment relationship, where it would remove the conditions under which the disclosures we study occur.

This study was not subject to IRB review as it involves secondary analysis of publicly available data with no interaction with human subjects.

\appendix

\section{Screening Annotation Guideline}
\label{app:screen}

Posts were annotated against a fine-grained scheme
capturing distinct forms of work-related distress,
then collapsed to a binary label for the screening
classifier in Section~\ref{sec:corpus}. Any post
assigned one of categories 1 through 10 was
labeled in scope. Category 0 captures posts
unrelated to work stress entirely, such as pure
technical questions, news and memes. The breadth
is deliberate. It establishes a high-recall
screening stage so that posts expressing burnout
indirectly are not excluded before the narrower
BAT annotation is applied.

\begin{table}[h]
\centering
\small
\setlength{\tabcolsep}{4pt}
\caption{Screening categories. Categories 1 to 10
collapse to in scope.}
\label{tab:screening}
\begin{tabular}{cp{6.1cm}}
\toprule
\# & Description \\
\midrule
0  & Not about work stress or burnout \\
1  & Work-related burnout, chronic stress,
workload, job demands \\
2  & Ambiguous work stress; frustration, mental
exhaustion or depressive affect \\
3  & Imposter syndrome at work; not knowing
something work-related \\
4  & PTSD, mental health disorder or illness \\
5  & De-stressing, stress relief, avoiding
overwhelm \\
6  & Work-life balance, work-health balance,
workload balance \\
7  & Job search or interview difficulty, job
change, ethical or trust issues, toxic environment
\\
8  & Anxiety \\
9  & Staying motivated, learning new things \\
10 & Sleep disturbance, emotional numbness \\
\bottomrule
\end{tabular}
\end{table}

\section{BAT Construct Annotation Prompt}
\label{app:bat}

The prompt below was given to Kimi K2.5 for every
screened post. The model returned one judgment per
symptom, and for a positive judgment the exact
phrase that triggered it.

\begin{promptbox}{General
instructions}{colframe=black!60}
You are a researcher applying the Burnout
Assessment
Tool (BAT) to Reddit text. Decide YES or NO for
each of
four symptoms. Count a burnout or stress signal
written
in any tense, including anticipated stress, past
stress,
or present difficulty with sleep or work-life
balance.

- Read the text cold. Assume nothing about whether
  burnout is present.
- This is a sensitivity-first task. When in doubt,
lean
  YES. Missing a real signal is worse than
  flagging a
  stress-adjacent one.
- A post asking others about their experience is
not the
  same as expressing it yourself.
- Reddit language rarely uses clinical terms. Look
for
  the meaning, not the exact words.
- The word "burnout" alone with no other signal is
NO.
  Any supporting signal alongside it is YES.
\end{promptbox}

\begin{promptbox}{Exhaustion}{colframe=blue!50!black}
Energy loss from work, physical or mental,
including
sustained overload that implies depletion even
without
the word "drained".

YES if: explicit depletion ("drained", "nothing
left",
  "running on empty"); sustained overload
  described over
  weeks or months; physical or health
  deterioration
  attributed to work; persistent misery tied to
  the job;
  no energy to begin work.

NO if:  a single bad day with no sustained
element;
  asking others about exhaustion rather than
  expressing
  it; boredom or dissatisfaction with no energy or
  health cost described.
\end{promptbox}

\begin{promptbox}{Emotional Impairment}{colframe=teal!50!black}
Intense, persistent or disproportionate emotional
reactions tied to work. Does not require explicit
loss
of control. Strong sustained negative emotion
qualifies.

YES if: strong aversion toward the work situation;
  repeated or stacked emotional signals in the
  same post;
  snapping, unexpected crying or overreacting at
  work;
  persistent irritability beyond a single
  incident;
  feeling unable to control emotions at work.

NO if:  a single proportionate frustration
mentioned
  once, calmly; mild annoyance without intensity
  or
  repetition; asking others about frustration
  rather
  than expressing it.
\end{promptbox}

\begin{promptbox}{Cognitive Impairment}{colframe=orange!55!black}
Difficulty with memory, focus or decisions at
work,
including feeling cognitively overwhelmed by the
volume,
complexity or pace of the job.

YES if: overwhelmed by cognitive demands such as
alert
  volume or task complexity, beyond normal
  new-role
  adjustment; brain fog, forgetting procedures,
  trouble
  concentrating; indecision on normally easy
  decisions;
  difficulty keeping up with job demands.

NO if:  a new hire describing normal learning
difficulty
  with no distress; asking others about cognitive
  difficulty rather than expressing it; general
  confusion
  with no work-specific symptom.
\end{promptbox}

\begin{promptbox}{Mental Distance}{colframe=purple!55!black}
Persistent psychological withdrawal: indifference,
cynicism, aversion, or working on autopilot.

YES if: explicit loss of meaning or interest
("what is
  the point", "I do not care anymore"); going
  through
  the motions or autopilot, described personally;
  active
  avoidance of work tasks or colleagues;
  persistent
  cynical or resentful tone; persistent dread of
  work.

NO if:  asking others about engagement rather than
  expressing own detachment; a single bad day or
  one-off
  complaint; considering a career change out of
  ambition
  or curiosity rather than withdrawal; mild
  boredom
  mentioned once without sustained withdrawal; a
  stated
  plan to leave with no accompanying indifference,
  cynicism or withdrawal.
\end{promptbox}

\begin{promptbox}{Output format}{colframe=black!60}
Respond in JSON only, with no markdown fences and
no
text outside the object. For each symptom return a
binary judgment and a reasoning field. For YES,
the
reasoning field is the exact phrase from the post
that
triggered the judgment. For NO, it is a
one-sentence
explanation of why the text did not meet the
threshold.
\end{promptbox}

\section{Exit Intention Prompt}
\label{app:exit}

This pass reads only for directional language about leaving. It makes no reference to the burnout symptoms, so the two annotations stay independent of one another.

\begin{promptbox}{Exit intention classification}{colframe=red!45!black}
You are labelling Reddit posts written by security
practitioners. Decide whether the author expresses
an
intention to leave their job or the field. Read
only for
language about leaving. Do not infer intention
from how
difficult the situation sounds.

Return exactly one of three labels.

exit_explicit
  A decision stated or an action already taken.
  Examples of the form: "I put my notice in",
  "I start somewhere new in March", "I am done
  with
  this field".

exit_contemplating
  Weighing leaving, without having decided.
  Examples:
  "I have started looking", "thinking about
  getting
  out", "not sure how much longer I can do this".

no_exit
  Difficulty, frustration or exhaustion with no
  directional language about leaving. Venting
  alone is
  not exit intention. Asking how others cope is
  not
  exit intention.

Rules
- Wanting a different role at the same employer
counts
  as exit_contemplating only if framed as leaving
  the
  current job, not as ordinary internal movement.
- A hypothetical about the field in general is not
exit
  intention unless applied to the author.
- Past departures from previous jobs are not
current
  exit intention.

Return JSON only:
  {"label": ..., "confidence": 0.0 to 1.0,
   "evidence": "<exact phrase from the post>",
   "reasoning": "<one sentence>"}
\end{promptbox}

\begin{table}[h!]
\centering
\small
\setlength{\tabcolsep}{4pt}
\caption{Latent class estimates of each rater's sensitivity and specificity (\%), fitted to all 100 validation posts including those where the coders disagreed.}
\label{tab:latentclass}
\begin{tabular}{lcccccc}
\toprule
& \multicolumn{2}{c}{Coder 1} & \multicolumn{2}{c}{Coder 2} & \multicolumn{2}{c}{Kimi K2.5} \\
\cmidrule(lr){2-3} \cmidrule(lr){4-5} \cmidrule(lr){6-7}
Symptom & Sens & Spec & Sens & Spec & Sens & Spec \\
\midrule
Exhaustion            & 95 & 94 &  98 &  92 & 77 & 87 \\
Emotional impairment  & 69 & 89 &  94 & 100 & 85 & 78 \\
Cognitive impairment  & 67 & 97 &  40 &  98 & 75 & 93 \\
Mental distance       & 75 & 98 & 100 &  78 & 87 & 88 \\
\bottomrule
\end{tabular}
\end{table}

\section{Concept Label Criteria}
\label{app:labels}

Table~\ref{tab:labels} gives the ten labels used to describe what symptomatic posts are about, and Table~\ref{tab:replylabels} the fourteen used to describe what replies do. Every label was scored independently on each document, as a binary judgment with no rationale requested. The induced labels were written by the concept induction step and left unedited. The labels drawn from prior work were written by us only after the induced set was frozen, so that induction was not steered toward categories we already had.

\begin{table*}[t]
\centering
\small
\caption{The ten content labels used to describe what symptomatic posts are about. The first four were induced from the posts themselves, with no categories supplied in advance. The remaining six come from prior turnover research and were written only after the induced set was frozen.}
\label{tab:labels}
\begin{tabular}{@{}ll@{}}
\toprule
Label & What it marks \\
\midrule
\multicolumn{2}{@{}l}{\textit{Induced from the corpus}}\\
\quad Workplace career strain & Unsustainable demands, declining wellbeing, or thoughts of changing job or field \\
\quad IT operations dysfunction & Understaffing, excessive technical responsibility, weak processes, or unclear priorities \\
\quad Security governance resistance & Management ignoring, delaying or overriding controls, compliance or risk findings \\
\quad Learning confidence struggles & Feeling unqualified or overwhelmed while developing skills or handling unfamiliar work \\
\addlinespace[3pt]
\multicolumn{2}{@{}l}{\textit{From prior turnover research}}\\
\quad Management failure & Poor decisions, or absent support, from management \\
\quad Resource inadequacy & Too little budget, tooling or headcount for the work being asked \\
\quad Technical disillusionment & Loss of belief in the value or effectiveness of the work itself \\
\quad Toxic culture & Hostile or corrosive relationships at work \\
\quad Compensation & Pay judged inadequate for the responsibility carried \\
\quad Value misalignment & Mismatch between the practitioner's professional values and the employer's priorities \\
\bottomrule
\end{tabular}
\end{table*}

\begin{table*}[t]
\centering
\small
\caption{The fourteen labels used to describe what replies do. The eight on the left were induced from the replies themselves. The six on the right are established categories of social support.}
\label{tab:replylabels}
\begin{tabular}{@{}ll@{}}
\toprule
Label & What it marks \\
\midrule
\multicolumn{2}{@{}l}{\textit{Induced from the replies}}\\
\quad Career planning advice & Advice on changing role, employer, specialism or career direction \\
\quad Workplace problem guidance & Advice on handling the specific situation the poster described \\
\quad Practical advice & Concrete technical or procedural steps to take \\
\quad Emotional support & Care, sympathy, reassurance or encouragement toward the poster \\
\quad Personal relating & A comparable experience of the replier's own, rather than a solution \\
\quad Support connections & Pointing the poster toward another person, community or resource \\
\quad Critical pushback & Challenging, minimising or dismissing the poster's account \\
\quad Discussion direction & Redirecting the thread, asking for clarification, or reframing the question \\
\addlinespace[3pt]
\multicolumn{2}{@{}l}{\textit{From the social support literature}}\\
\quad Informational support & Advice, guidance, instruction or factual help \\
\quad Emotional support & Care, sympathy, reassurance or encouragement \\
\quad Esteem support & Affirming the poster's worth, competence or judgment \\
\quad Tangible support & A concrete offer of help, resources or action \\
\quad Network support & Signalling shared experience or belonging \\
\quad Unsupportive response & Dismissal, minimisation, criticism or hostility \\
\bottomrule
\end{tabular}
\end{table*}

\end{document}